\documentclass[twocolumn,showpacs,preprintnumbers,amsmath,amssymb]{revtex4}
\usepackage{graphicx}% Include figure files
\usepackage{dcolumn}% Align table columns on decimal point
\usepackage{bm}% bold math

\begin{document}

\input epsf.sty

%\draft

%\twocolumn[\hsize\textwidth\columnwidth\hsize\csname %
%@twocolumnfalse\endcsname

\title{Phase separation in the vicinity of
``quantum critical'' doping concentration: implications for high temperature superconductors}

\author{B. V. Fine$^1$}
\email{bfine@utk.edu}

\author{T. Egami$^{1,2,3}$}

\email{egami@utk.edu}

\affiliation{
$^1$ Department of Physics and Astronomy, University of Tennessee, Knoxville, TN 37996, USA \\
$^2$ Department of Materials Science and Engineering, University of Tennessee, Knoxville, Tennessee 37996, USA \\
$^3$ Oak Ridge National Laboratory, Oak Ridge, Tennessee 37831, USA}

\date{July 25, 2007}

\begin{abstract}
A general quantitative measure of the tendency towards phase separation is introduced for systems exhibiting phase transitions or crossovers controlled by charge carrier concentration. This measure is devised for the situations when the quantitative knowledge of various contributions to free energy is incomplete, and is applied to evaluate the chances of electronic phase separation associated with the onset of antiferromagnetic correlations in high-temperature cuprate superconductors. The experimental phenomenology of lanthanum- and yittrium-based cuprates was used as input to this analysis. It is also pointed out that Coulomb repulsion between charge carriers separated by the distances of 1-3 lattice periods strengthens the tendency towards phase separation by accelerating the decay of antiferromagnetic correlations with doping. Overall, the present analysis indicates that cuprates are realistically close to the threshold of phase separation --- nanoscale limited or even macroscopic with charge density varying between adjacent crystal planes.
\end{abstract}

%\vskip 2cm\centerline{PLEASE DO NOT DISTRIBUTE.}
%\vskip 2cm

\maketitle

%\narrowtext
%\pagebreak

\section{Introduction}

Phase transitions or sharp crossovers induced by continuous change in chemical composition have natural tendency to cause phase separation. Phase separation occurs, when the free energy of the most stable homogeneous state has concave dependence on the concentration of particles (i.e. this dependence is curved or peaked upward). The concave behavior is particularly likely in the course of a phase transition or a crossover, because the free energy switches between two different dependences characteristic of adjacent phases.
The steeper is this switching, the likelier is the onset of concave behavior. In the case of electronic subsystems in solids, a strong Coulomb repulsion raises the threshold of phase separation and also, if the homogeneous electronic state becomes unstable,  limits the size of single-phase regions to the  nanometer scale along at least one of the spatial directions. In the present work, we introduce a quantitative measure of the above tendency towards phase separation in the vicinity of phase transitions and crossovers for the situations when the knowledge of various contributions to free energy is incomplete. We justify this measure in a very general context, and then use it to analyze the chances of electronic phase separation in high temperature cuprate superconductors.

Macroscopic phase separation in cuprates had been observed in LaCuO$_{4+\delta}$\cite{Radaelli-94}, where it occurs because of the high mobility of intercalated oxygen that maintains charge neutrality within each phase. In other cuprate families, where dopant ions are immobile, the experimental situation is complicated by the possibility that phase separation may be limited by Coulomb interaction to the difficult-to-access scale of a few nanometers, where inhomogeneities can also fluctuate in time. The direct experimental evidence for static nanoscale modulations has been reported so far only for a rather limited subset of cuprates\cite{Tranquada-95,Fujita-04,Hoffman-02,Hoffman-02A,Howald-03,Vershinin-04,Egami-06}.

Theoretically, the general tendency towards phase separation in cuprates and cuprate-related models is extensively discussed in the literature\cite{Emery-90,Grilli-91,Sigmund-92,Emery-93,Dagotto-94,Nagaev-95,Hellberg-97,Markiewicz-97,Becca-00,Lorenzana-01,Kotliar-02,Carlson-03,Goodenough-03,Capone-04,Ivanov-04,Aichhorn-05,Aichhorn-06,Lugas-06,Macridin-06,Ortix-06,Eckstein-07,Kocharian-07},  but any conclusion about the actual presence or absence of phase separation suffers from uncertainties associated with the gross simplifications in the models. The important advantage of the phase separation measure introduced in this work is that it is directly applicable to real materials. At the same time, our analysis also provides a useful perspective to the results of various model studies.

Cuprates can exhibit phase separation for two related reasons: the onset of antiferromagnetic (AF) correlations and the onset of Mott insulating gap. In this work, we focus on the former and arrive at the conclusion that cuprates are very close to the phase separation threshold and can realistically phase separate. The The experimental results on the doping dependence of the overall intensity of antiferromagnetic fluctuations in YBa$_2$Cu$_3$O$_{6+x}$\cite{Bourges-00} and La$_{2-x}$Sr$_x$CuO$_4$\cite{Wakimoto-07} constitute important input into our analysis. We also propose how the phenomenology of phase separation in LaCuO$_{4+\delta}$\cite{Radaelli-94}  can be used to obtain information about the properties of ``generic'' cuprates. Finally we point out that Coulomb repulsion between charge carriers separated by the distances of 1-3 lattice periods {\it strengthens} the tendency towards phase separation by accelerating the decay of AF correlations with doping.

\section{Phase transitions and phase separation}

\subsection{Free energy}
\label{free}

\begin{figure} \setlength{\unitlength}{0.1cm}
%=======================================================================

\begin{picture}(100, 60)
{ 
\put(10, 0){ \epsfxsize= 2.3in \epsfbox{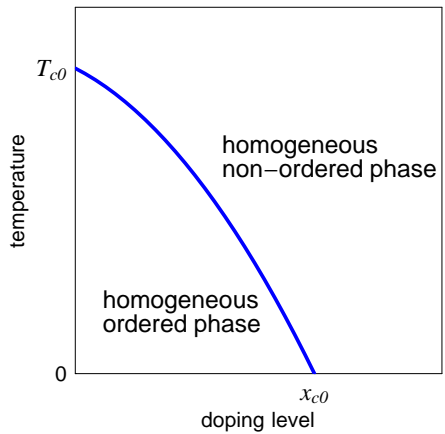} }
}
\end{picture} 
%============== 
\caption{(Color online) Temperature-vs.-concentration phase diagram for homogeneous phases. The solid line can represent a phase transition of any order or a crossover.
} 
\label{fig-Tcx} 
\end{figure}

We analyze the dependence of free energy of energetically most stable homogeneous state as a function of charge carrier concentration. Negative second derivative of this dependence (equivalent to negative curvature or negative inverse compressibility) will be considered as a sign of phase separation. Motivated by the physics of cuprates, we consider the ``homogeneous'' temperature-vs.-concentration phase diagram looking as shown in Fig.~\ref{fig-Tcx}.

%%%%%%%%%%%%%%%%%%%%%%%%%%%%%%%%%%%%%%%%%%%%%%%%%%%%%%%%%%%%%%%%%%%

% %%%%%%%%%%%%%%%%%%%%%%%%%%%%%%%%%%%%%%%%%%%%%%%%%%%%%%%%%%%%%%%%%

\begin{figure}
\setlength{\unitlength}{0.1cm}
%=======================================================================
\begin{picture}(100, 71)
{
\put(4, 0){
\epsfxsize=3in
\epsfbox{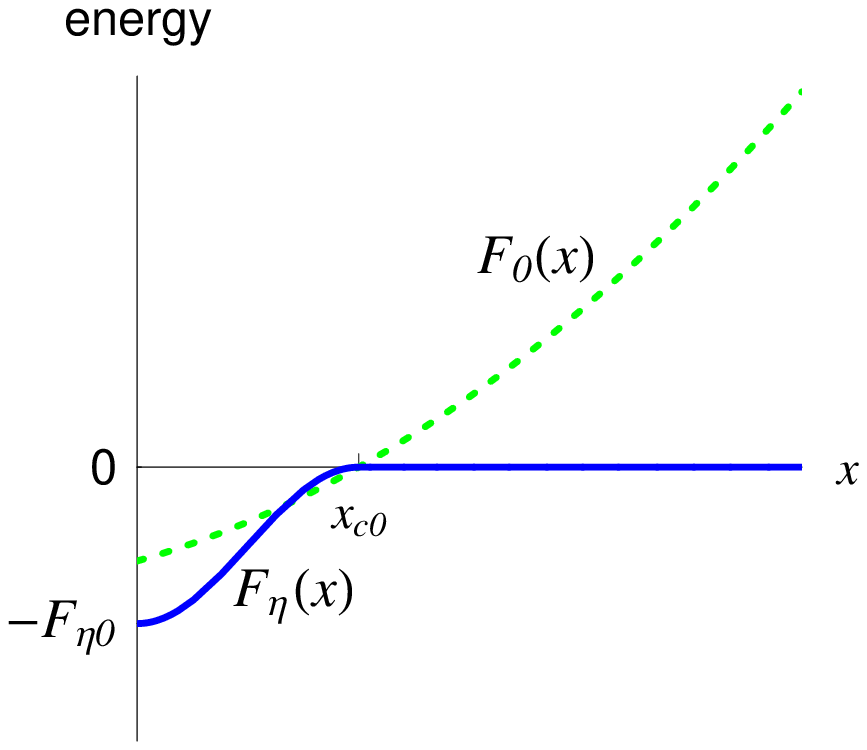} }
\put(5,58){\textsf{\Large (a)}}
}
\end{picture} 
\begin{picture}(100, 75)
{
\put(9, 0){
\epsfxsize=3in
\epsfbox{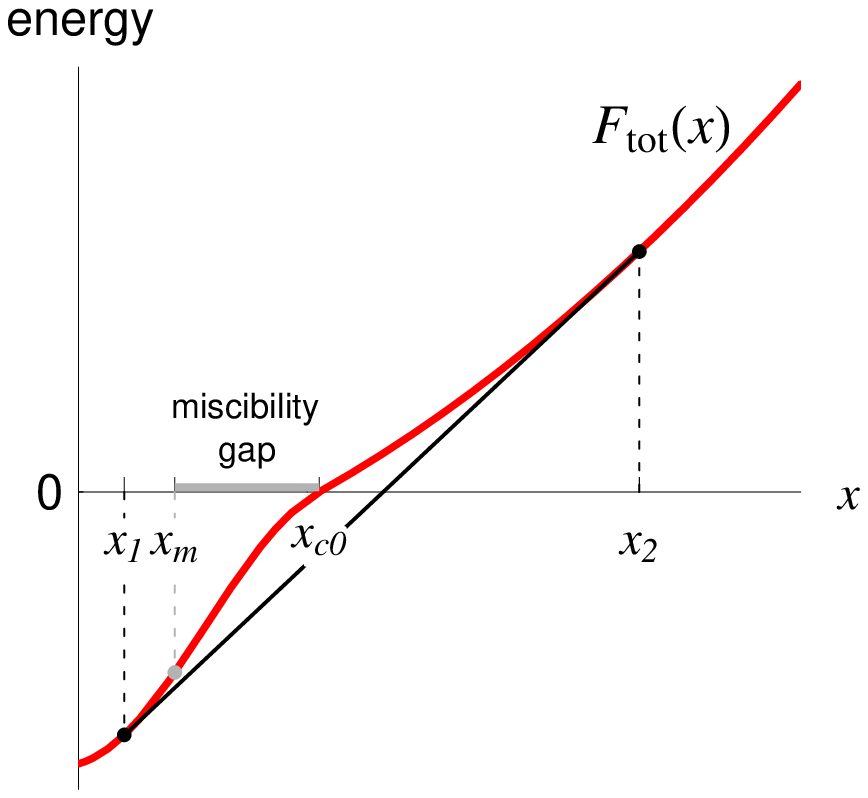} }
\put(5,60){\textsf{\Large (b)}}
}
\end{picture} 
%============== 
\caption{(Color online) Sketches of free energy as a function of charge carrier concentration for a phase transition accompanied by phase separation. (a) Free energy of non-ordered state $F_0(x)$, and free energy gain $F_{\eta}(x)$ due to the phase transition at $x_{c0}$. (b) Total free energy $F_{\hbox{\small tot}}(x)$ (thick line) equal to the sum of $F_0(x)$ and $F_{\eta}(x)$ sketched in (a), and the Maxwell construction.
}
\label{fig-F} 
\end{figure}

The total free energy of the system $F_{\hbox{\small tot}}$ as a function of dimensionless charge carrier concentration $x$ can be decomposed as shown in Fig.~\ref{fig-F}, namely:
\begin{equation}
F_{\hbox{\small tot}} (x) = F_0 (x) + F_{\eta}(x),
\label{Ftot}
\end{equation} 
where $F_0(x)$ is the free energy of a non-ordered state (e.g. non-magnetic Fermi liquid), and  $F_{\eta}(x)$ is the part of free energy associated with the onset of a more ordered state below the transition or crossover at $x=x_{c0}$ (e.g. AF Neel phase or some kind of electronic liquid with strong AF correlations like resonating valence bond (RVB) state\cite{Anderson-87}).

Let us now introduce the Maxwell construction by assuming that we are dealing with an idealized neutral system, where particles separate into two macroscopic regions: fraction $f_1$ of them forms regions with the concentration $x_1$, while fraction $f_2$ goes into the concentration $x_2$. The free energy $F_{\hbox{\small sep}}$ of this phase separated mix is given by equation
\begin{equation}
F_{\hbox{\small sep}} = f_1 F_{\hbox{\small tot}}(x_1) +  f_2 F_{\hbox{\small tot}}(x_2),
\label{Ftilde}
\end{equation}
with two constraints
\begin{equation}
f_1 +  f_2 = 1,
\label{fnorm}
\end{equation}
\begin{equation}
f_1 x_1 +  f_2 x_2 = x_{\hbox{\small av}},
\label{nnorm}
\end{equation} 
where $x_{\hbox{\small av}}$ is the average concentration set by the doping level.
The minimization of $F_{\hbox{\small sep}}$ with respect to $f_1$, $f_2$, $x_1$ and $x_2$ under conditions (\ref{fnorm}, \ref{nnorm}) gives:
\begin{equation}
\left. {d F_{\hbox{\small tot}} \over dx} \right|_{x = x_1} =
\left. {d F_{\hbox{\small tot}} \over dx} \right|_{x = x_2} = {F_{\hbox{\small tot}}(x_2) - F_{\hbox{\small tot}}(x_1) \over  x_2 - x_1}
\label{cond1}
\end{equation}
The derivatives ${d F_{\hbox{\small tot}} \over dx}$ are, of course, the chemical potentials of the two phases.

%%%%%%%%%%%%%%%%%%%%%%%%%%%%%%%%%%%%%%%%%%%%%%%%%%%%%%%%%%%%%%%%%%%

% %%%%%%%%%%%%%%%%%%%%%%%%%%%%%%%%%%%%%%%%%%%%%%%%%%%%%%%%%%%%%%%%%

The above condition has transparent geometrical interpretation --- Maxwell construction --- illustrated in Fig.~\ref{fig-F}(b):
The line connecting points $(x_1, F_{\hbox{\small tot}}(x_1))$ and $(x_2, F_{\hbox{\small tot}}(x_2))$ should be tangential to the curve $F_{\hbox{\small tot}}(x)$ at the both points. The energy of the phase separated state is then a point on this line at $x=x_{\hbox{\small av}}$. The necessary condition for Eq.(\ref{cond1}) to correspond to a minimum rather than a maximum is to have a region of negative curvature somewhere between $x_1$ and $x_2$.

If $F_{\hbox{\small tot}}(x)$ and its first derivative are continuous everywhere, then there exist three different regions of carrier concentrations:

1) Miscibility gap $x_m < x < x_{c0}$, where the curvature is negative, and hence, the uniform composition locally unstable towards spinodal decomposition.

2) Metastable regions $x_1 < x < x_m$ and $x_{c0} < x < x_2$, where the homogeneous state is locally stable due to the positiveness of the curvature, but globally unstable, because
the fully separated inhomogeneous state has lower energy. 

3) Stable regions $x < x_1$ and $x_2 < x$.

If $F_1(x)$ exhibits a cusp\cite{cusp}, then one of the above regions can shrink to a point.

\

Now we return to the discussion of real systems described by free energy (\ref{Ftot}). We assume that the non-ordered state does not phase separate on its own, and, therefore, the curvature of $F_0(x)$, which we define as
\begin{equation}
K_0 \equiv {1 \over 2} {d^2 F_0 \over dx^2},
\label{K0Def}
\end{equation}
is positive everywhere. [This and other curvatures introduced below are implied to be $x$-dependent.] The onset of the more ordered state should necessarily lower the energy of the system, which, as obvious from Fig.~\ref{fig-F}(a), implies that $F_{\eta}(x)$ should have either intervals of finite negative curvature, or a point of infinite negative curvature (upward cusp).
 We define the ``negative'' curvature of $F_{\eta}(x)$ as
\begin{equation}
K_{\eta} \equiv - {1 \over 2} {d^2 F_{\eta} \over dx^2}.
\label{KetaDef}
\end{equation}
The curvature of the total homogeneous free energy can now be expressed as
\begin{equation}
K_{\hbox{\small tot}} \equiv {1 \over 2} {d^2 F_{\hbox{\small tot}} \over dx^2} = K_0 - K_{\eta}  .
\label{Ktot}
\end{equation}

A negative value of $K_{\hbox{\small tot}}$ is necessary but not sufficient condition for the onset of phase separation. Once an excess charge carrier concentration $\Delta x$ starts building in some part of the sample, there is an additional energy cost $\Delta F (\Delta x)$ associated with the Coulomb repulsion of uncompensated charge, strain energy and gradient energy. When dopant atoms are immobile, we expect the main additional contribution to come from Coulomb energy and denote the positive curvature associated with this contribution as
\begin{equation}
K_{\hbox{\small Coul}} \approx {1 \over 2} {d^2 \Delta F \over d\Delta x^2}.
\label{KCoulDef}
\end{equation}
The condition for the onset of phase separation then becomes:
\begin{equation}
K_{\eta} \geq K_0 + K_{\hbox{\small Coul}}  ,
\label{Ktot}
\end{equation}
and the Maxwell construction bounds on the miscibility gap and the metastable regions should also be modified accordingly.

In subsection~\ref{rigorous} we quantify how large $K_{\eta}$ can be for the situations when only the values of $F_{\eta}(0)$ and $x_{c0}$ are known, with
or without additional knowledge of  ${ dF_{\eta}(x) \over dx}$ at $x=x_{c0}$ and/or $x = 0$.

\subsection{Different scenarios of phase separation}
\label{routes}

The fact that the system phase separates is sometimes perceived as implying first order phase transition. However, in the case of volume constrained systems, e.g. electrons on the lattice, the relation between phase separation and the order of phase transition becomes less straightforward.
In contrast to the more familiar constraint of constant pressure, which allows the phase coexistence only at a fixed temperature of a first order phase transition, the constraint of constant volume allows the coexistence of two phases in a range of temperatures\cite{Landau-80}. Such a property, in turn, defeats the classification of the resulting behavior in terms of the jumps of thermodynamic derivatives, because, when the temperature of the system decreases past the phase separation threshold, the volume fraction of the second phase increases gradually, which normally implies no jump in entropy and no latent heat, but leads to a jump in the specific heat like in a second order phase transition. Yet, at the phase separation temperature, the homogeneous state usually remains metastable, which means, that, in order to become phase separated, the system has to overcome a nucleation threshold.

\begin{figure} \setlength{\unitlength}{0.1cm}
%=======================================================================

\begin{picture}(100, 62) 
{ 
\put(5, 0){ \epsfxsize= 3in \epsfbox{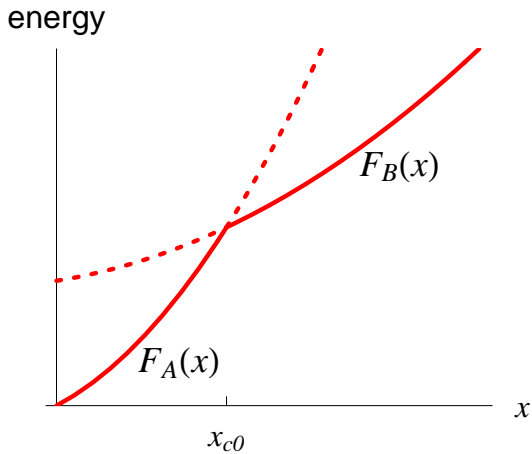} }
}
\end{picture} 
%============== 
\caption{ (Color online) Sketch of a first order phase transition. The system switches from a potential minimum $F_A(x)$ associated with state $A$ to the potential minimum $F_B(x)$ associated with state $B$. This transition necessarily leads to an upward cusp.
} 
\label{fig-cusp} 
\end{figure}

\begin{figure} \setlength{\unitlength}{0.1cm}
%=======================================================================

\begin{picture}(100, 68) 
{ 
\put(5, 0){ \epsfxsize= 3in \epsfbox{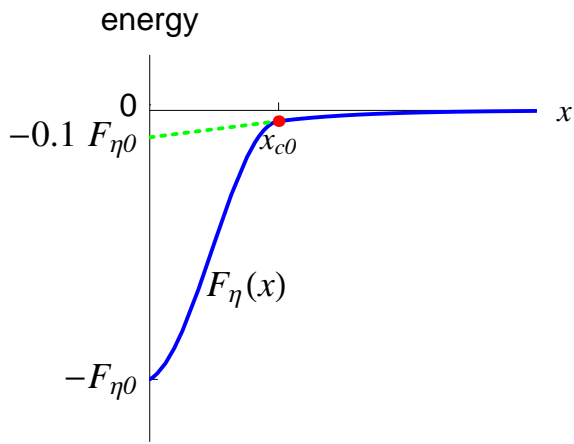} }
}
\end{picture} 
%============== 
\caption{ (Color online) Sketch of free energy gain $F_{\eta}(x)$ associated with a crossover (solid line). Dashed line illustrates the graphical meaning of convention(\ref{Fc}) for fixing the location of the crossover point $x_{c0}$.
} 
\label{fig-cross} 
\end{figure}

In our classification, we distinguish between the types of ``would-be" homogeneous phase transitions, which are behind the onset of phase separation. Eventual phase separation is compatible with the possibilities that the line in  Fig.~\ref{fig-Tcx} represents a phase transition of first order (Fig.~\ref{fig-cusp}) --- always exhibiting a cusp, second or higher orders (Fig.~\ref{fig-F}) --- with or without cusp dependently on the critical exponents, or it may also represent a crossover (Fig.~\ref{fig-cross}). These are distinct scenarios of phase separation with distinct properties. Before proceeding further, the readers may choose to read the discussion of these scenarios given in Appendix~\ref{classification}.  The part of this Appendix on the second order phase transitions exemplifies the general constraint on the negative curvature to be derived in the next subsection.

%%%%%%%%%%%%%%%%%%%%%%%%%%%%%%%%%%%%%%%%%%%%%%%%%%%%%%%%%%%%%%%%%%%

% %%%%%%%%%%%%%%%%%%%%%%%%%%%%%%%%%%%%%%%%%%%%%%%%%%%%%%%%%%%%%%%%%

%%%%%%%%%%%%%%%%%%%%%%%%%%%%%%%%%%%%%%%%%%%%%%%%%%%%%%%%%%%%%%%%%%%

% %%%%%%%%%%%%%%%%%%%%%%%%%%%%%%%%%%%%%%%%%%%%%%%%%%%%%%%%%%%%%%%%%

\subsection{Rigorous constraints on the negative curvature}
\label{rigorous}

In this subsection, we consider the following problem and its generalizations. 

{\it Problem A} --- Phase transition: Let us assume, that we can isolate certain contribution $F_{\eta}(x)$ to the free energy of the system as a function of charge carrier concentration $x$. About this contribution, we only know that, at $x=0$,   $F_{\eta}(0) = -F_{\eta 0} < 0$ and, for carrier concentrations greater than some given positive value $x_{c0}$, $F_{\eta}(x) = 0$. [The sketch for $F_{\eta}(x)$ is given in Fig.~\ref{fig-F}(a).]
The latter condition implies that $F_{\eta}(x_{c0}) = 0$ and
$F_{\eta}^{\prime}(x_{c0}+0) = 0$. (Here and everywhere, $F^{\prime}(x)$ refers to the first derivative with respect to $x$, and $F^{\prime \prime}(x)$ refers to the second derivative.)
Any path leading from $F_{\eta}(x_{c0})$ to $F_{\eta}(0)$ should have a region of negative curvature (or a single point with infinite negative curvature).  Let us further assume that, for every possible path $F_{\eta}(x)$ satisfying the above conditions, we can find the value of the maximum negative curvature (max$[-{1 \over 2}F_{\eta}^{\prime \prime} (x)]$) in the interval $[0,x_{c0}]$. The question we ask is: What is the {\it minimum} possible value of max$[-{1 \over 2}F_{\eta}^{\prime \prime} (x)]$?  We denote this value as $K_{\eta M}$.

\begin{figure} \setlength{\unitlength}{0.1cm}
%=======================================================================

\begin{picture}(100, 71) 
{ 
\put(5, 0){ \epsfxsize= 3in \epsfbox{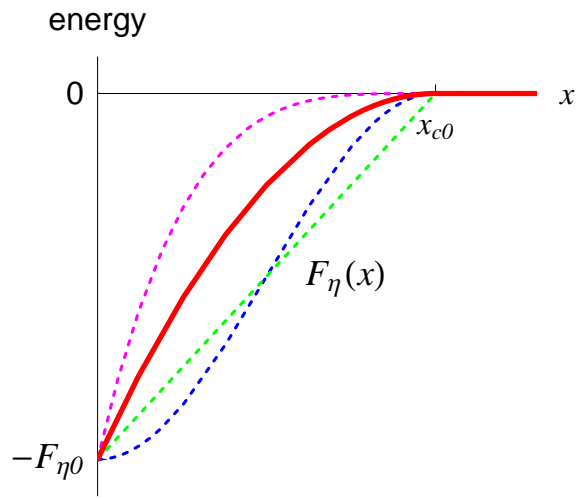} }
}
\end{picture} 
%============== 
\caption{ (Color online)  Parabola (solid line) minimizing the value of the maximum negative curvature for the boundary conditions of Problem A, and some
obvious alternatives to this parabola (dashed lines). The straight-line alternative ends in an upward cusp at $x=x_{c0}$, which means a cusp with infinite negative curvature. Two other alternatives have intervals with finite negative curvatures larger than that of parabola.
} 
\label{fig-alt} 
\end{figure}

A rigorous solution of this problem is presented in Appendix~\ref{constraints}. The result is
\begin{equation}
K_{\eta M} =  {F_{\eta 0} \over x_{c0}^2},
\label{minKmax1}
\end{equation}
which is equivalent to the expression (\ref{curv}) obtained in Appendix~\ref{classification} in the framework of Landau expansion for a second order phase transition. Curvature (\ref{minKmax1}) corresponds to the parabola shown in Fig.\ref{fig-alt}. Other sketches in the same figure represent obvious alternatives to the parabolic dependence. These alternatives exhibit either regions of larger negative curvature, or a point of infinitely negative curvature (cusp) at $x = x_{c0}$ --- both are the possibilities that are stronger in favor of eventual phase separation.

%%%%%%%%%%%%%%%%%%%%%%%%%%%%%%%%%%%%%%%%%%%%%%%%%%%%%%%%%%%%%%%%%%%

% %%%%%%%%%%%%%%%%%%%%%%%%%%%%%%%%%%%%%%%%%%%%%%%%%%%%%%%%%%%%%%%%%

Constraint (\ref{minKmax1}) will play the central role in our subsequent analysis of phase transitions and crossovers.

\

Problem A can be straightforwardly generalized to the following one.

{\it Problem A1} --- Crossover: Assume that
\mbox{$F_{\eta}(x_{c0}) = - F_{\eta C}$}
and
\mbox{$F_{\eta}^{\prime}(x_{c0}) = F_{\eta C}^{\prime} < (F_{\eta 0}-F_{\eta C})/x_{c0}$},
where $F_{\eta C}$ and $F_{\eta C}^{\prime}$ are two new constants.
All other conditions are the same as in Problem A.

The solution of Problem A1 is (see Appendix~\ref{constraints})
\begin{equation}
K_{\eta M} =  {F_{\eta 0} - F_{\eta C} - F_{\eta C}^{\prime} x_{c0} \over x_{c0}^2}.
\label{minKmax2}
\end{equation}
In the case $F^{\prime}_{\eta C} \geq (F_{\eta 0}-F_{\eta C})/x_{c0}$, there are paths between $x=0$ and $x=x_{c0}$, which do not exhibit negative curvature.

Expression (\ref{minKmax2}) is useful, when one needs to estimate the error associated with applying the basic expression (\ref{minKmax1}) to a crossover, in which case the choice of $x_{c0}$ may be somewhat arbitrary. That choice, however, can be made unambiguous by requiring, e.g., that
\begin{equation}
F_{\eta C} + F^{\prime}_{\eta C} x_{c0} = 0.1 F_{\eta 0}.
\label{Fc}
\end{equation}
In this case, the error due to the application of expression (\ref{minKmax1}) is limited to 10 per cent.

\

In some physical situations, the value of $F_{\eta}^{\prime}(0)$ may also be known. This knowledge can then be exploited as follows.

{\it Problem A2}:  $F_{\eta}^{\prime}(0) = F^{\prime}_{\eta 0}$, where
$F^{\prime}_{\eta 0}$ is a new constant. All other conditions are the same as in Problem A1.

The solution of Problem A2 is given in the Appendix~\ref{constraints}. Here we only give the result:

1. If $ F^{\prime}_{\eta C} > (F_{\eta 0}-F_{\eta C})/x_{c0} > F^{\prime}_{\eta 0} $, then the paths between $x=0$ and $x=x_{c0}$
can have no negative curvature regions or points.

Otherwise:

2a. If 
\begin{equation}
F^{\prime}_{\eta 0} + F^{\prime}_{\eta C}  \leq {2 (F_{\eta 0}-F_{\eta C}) \over x_{c0}},
\label{Fp0Fpc}
\end{equation}
Then the minimum value of the maximum negative curvature is still given by Eq.(\ref{minKmax2}).

2b. If condition (\ref{Fp0Fpc}) is not fulfilled, then
\begin{equation}
K_{\eta M} =  {F_{\eta C} - F_{\eta 0} + F^{\prime}_{\eta 0} x_{c0} \over x_{c0}^2}.
\label{minKmax3}
\end{equation}

\section{Antiferromagnetic crossover in cuprates}
\label{crossover}

\subsection{Preliminary remarks}

In this part, we focus on the doping dependence of the energy of AF correlations in the cuprates, where the exchange coupling between Cu spins leads to establishing AF order at half-filling ($x=0$). As doping level $x$ increases, static AF order disappears through a second order phase transition. However, the magnetic energy change associated with this phase transition should not be large (in particular, in hole-doped cuprates), because the disappearance of the static order is due to the the loss of three-dimensional AF correlations between different CuO$_2$ planes. In the paramagnetic state, two-dimensional AF correlations should still carry large magnetic energy  in a broad parameter region near the transition line. The subsequent disappearance of magnetic energy associated with AF correlations can proceed either as a smooth crossover, or through one or several phase transitions involving possibly spin liquid states such as RVB.

In order to impose the negative curvature constraint (\ref{minKmax1}) on the doping evolution of magnetic energy, we do not need to know exactly what happens as the doping level increases. All we need to know is the easily accessible value of exchange energy at zero doping ($F_{\eta 0}$) and the value of critical doping $x_{c0}$, where condition (\ref{minKmax1}) is applicable.

\subsection{Decomposition of free energy}
\label{decomposition}

In Eq.(\ref{Ftot}), we assign $F_0(x)$ to be the lowest possible energy of a non-magnetic state at a given charge carrier concentration, and $F_{\eta}(x)$ to be the of the energy associated with the onset of AF correlations both static and dynamic. The latter includes both the AF exchange energy as such plus the change of kinetic, Coulomb and all other contributions to the total energy caused by AF correlations. With this definition,  once the AF correlations disappear, the non-magnetic state of the lowest energy is the actual physical state of the system. At the doping concentrations corresponding to non-zero AF correlations, the notion of the lowest energy non-magnetic state should be viewed as a variational ansatz, where the AF correlations are completely suppressed at and below typical exchange frequencies. It should be analogous to the paramagnetic state at temperatures much higher than exchange coupling.  At half doping ($x=0$), the optimal non-magnetic state should be a Mott insulator, where localized spins have completely random orientations.

\subsection{Estimates of positive and negative curvature contributions to free energy}

In this subsection we estimate curvatures $K_0$, $K_{\hbox{\small Coul}}$ and $K_{\eta}$ for the paramagnetic, Coulomb and antiferromagnetic contributions to free energy, respectively.
We are looking for conditions sufficient for phase separation as follows from inequality (\ref{Ktot}).  Therefore, at every step we try to overestimate the factors opposing phase separation, and underestimate factors leading to it.  [The constraint (\ref{minKmax1}) works this way.]

Everywhere below, the free energy and the corresponding curvature are evaluated per in-plane Cu atom. Occasionally, we interchange the terms ``doping" and ``charge carrier concentration".

\subsubsection{Paramagnetic curvature}

The positive curvature of $F_0(x)$ can be estimated as that of 2D Fermi-liquid
\begin{equation}
K_0 = {\pi \hbar^2 (1+f^s_0) \over 2 m^* a^2},
\label{K0A}
\end{equation}
where $m^*$ is the  effective mass and $f^s_0$ the Landau Fermi-liquid parameter\cite{Landau-57}.
Due to two-dimensionality, expression (\ref{K0A}) is helpfully independent of the Fermi-momentum and, hence, of the charge carrier concentration. Therefore, the above estimate would apply both to the conventional case, when all charge carriers form the Fermi sea, and, to an alternative case, when only doped charge carriers form a Fermi-liquid in the [remnants of] lower Mott-Hubbard band.

The effective mass can be estimated as
\begin{equation}
m^* = {3 \hbar^2 \gamma \over \pi k_B^2 a^2},
\label{mgamma}
\end{equation}
where $k_B$ is Boltzmann coefficient, and $\gamma$ is electronic specific heat coefficient per in-plane Cu atom for overdoped samples at temperatures high enough to destroy magnetic correlations. (Here, we ignore the low temperature pseudogap behavior in underdoped samples as the pseudogap itself can be the consequence of nanoscale-limited phase separation.)
For the estimate, we use $\gamma = 1 \ \hbox{mJ/gat.K$^2$}$ [converted per in-plane Cu atom as required by Eq.(\ref{mgamma})]. This value is consistent with experiments on La$_{2-x}$Sr$_x$CuO$_4$, YBa$_2$Cu$_3$O$_{6+x}$ and Bi$_2$Sr$_2$CaCu$_2$O$_{8+x}$\cite{Loram-01}. It leads to $m^* \approx 4 m_e$, where $m_e$ is the mass of free electron.   We are not aware of experiments, which would allow one to access the value of $f^s_0$ directly. It is, however, reasonable to assume that $f^s_0 \sim f^s_1 \approx 6$. Here, $f^s_1$  is another Fermi-liquid parameter determined by the effective mass according to relation $m^* = (1 + {1 \over 2} f^s_1) m_b \approx (1 + {1 \over 2} f^s_1) m_e$, where $m_b$ is the band mass. [In normal He-3 at zero applied pressure, $f^s_0 = 10.07$ and $f^s_1 = 6.04$\cite{Wheatley-75}.] As long as $f^s_1$ remains sufficiently large, assumption $f^s_0 \sim f^s_1$ guarantees that the resulting value of $K_0$ equals approximately twice the value for the free electron gas . Our final estimate in subsection~\ref{combine} will include allowance for a factor-of-two uncertainty in the above value.

The right-hand side of Eq.(\ref{K0A}) can be rewritten as $(1+f^s_0)/(2 \nu)$, where $\nu$ is the density of electronic states around the chemical potential. As such, this formula has a broader range of applicability: $\nu$ can be extracted from the specific heat measurements as $3 \gamma / (\pi k_B)^2$ irrespectively of whether the substance measured is a Fermi-liquid, and parameter $f^s_0$ characterizes the interaction between a newly added particle and the particles already present in the system.

The lowest energy non-magnetic state may or may not be a Fermi-liquid, especially around zero doping, where the onset of Mott insulating behavior (charge-transfer gap), the presence of random paramagnetic background, and the localization effects due to the random Coulomb potential of dopant atoms can drastically modify the behavior of charge carriers. The resulting metal-insulator transition is a fascinating subject on its own (see e.g. \cite{Kotliar-02,Markiewicz-04}). However, we feel, that, this transition (presumably around $x \approx 0.05$) can only further contribute to the tendency toward phase separation and, thereby, further strengthen the conclusions of this work.

\subsubsection{Coulomb curvature in the ``lasagna" scenario}
\label{Coulomb}

When the phase separation of 2D electronic systems is analyzed, one frequently mentioned picture is that of Coulomb-frustrated phase separation\cite{Emery-93,Nagaev-95,Lorenzana-01,Ortix-06}. Many researchers expect that Coulomb-frustrated in-plane phase separation can explain the formation of stripes, checkerboards or other nanoscale-limited inhomogeneous structures possibly existing in cuprates.

In the present work, however, we explore a different possibility, which we call the ``lasagna scenario''. According to this scenario, the three dimensional layered system phase separates into two-dimensional {\it macroscopic} regions of positive and negative charge lying on the top of each other in adjacent layers and having {\it equal in absolute value} and opposite in sign densities of uncompensated charge.
These regions effectively screen each other over the distances larger than the separation between layers. This kind of screening does not require a perfectly periodic charge order along the $c$-axis like $(+-+-+-+-)$, where $+$ or $-$ represents the sign of the charge. (Periodic arrangement would imply an observable doubling of the $c$-axis period.) A more disordered arrangement, where each positive layer has at least one adjacent negative layer and vice versa [e.g. $(+--+-++-+-)$] would be sufficient for the present scenario.

The Coulomb cost of this phase separation scenario can be estimated as that of Coulomb interaction between uncompensated charges within radius equal to the distance between the CuO$_2$ planes:
\begin{equation}
K_{\hbox{\small Coul}} = { 2 \pi e^2 R_s \over \epsilon a},
\label{KCoul} 
\end{equation}
where $e$ is the charge of electron, $R_s$ is the distance between CuO$_2$ planes, and $\epsilon$ is the dielectric constant estimated as 30.
Here we sidestep the discussion of multi-layered compounds and consider only single-layer cuprates.

The lasagna scenario is worth exploring seriously, since it may well be realized in cuprates, and this would influence strongly the interpretation of many experiments. At the same time, this scenario is easier to treat theoretically, since one does not need to deal with the gradient terms and the strongly fluctuational nature of nanoscale inhomogeneities. The Coulomb curvature (\ref{KCoul}) for our scenario is comparable with and can easily be lower than that of nanoscale limited phase separation. Absence of the nanoscale gradients is another energetic  advantage of this scenario. Where it may loose to the nanoscale-limited case is in the energy gained from phase separation. As follows from Maxwell construction [Fig.~\ref{fig-F}(b)], the charge densities of the oppositely charged regions in the phase separated state are, in general, not equal to each other. Imposing the equal charge condition means that the energy gain due to phase separation is less than optimal. Another extra cost of the present scenario is that of the suppressed interlayer hopping, which, however, should be small.

If the lasagna scenario can be proven to be energetically advantageous over the homogeneous  state, then it is guaranteed that the system is unstable towards phase separation --- either according to the lasagna scenario, or with in-plane nanoscale inhomogeneities, if they lower the energy even further or better accessible kinetically. The nanoscale-limited phase separation can also take place as a second stage, after charge density becomes different in adjacent planes according to the lasagna scenario.

\subsubsection{Antiferromagnetic curvature}

Here we apply constraint (\ref{minKmax1}). We estimate the value of $F_{\eta 0}$ in the approximation of staggered spin polarizations as
\begin{equation}
F_{\eta 0} = {1 \over 2} J N_{nn} s^2 = {1 \over 2} J,
\label{FetaJ}
\end{equation}
where $J$ is the exchange coupling constant, $N_{nn} = 4$ is the number of the nearest neighbors on the square lattice, and $s = 1/2$ is the value of Cu spin.

The constraint (\ref{minKmax1}) with definition (\ref{Fc}) for $x_{c0}$ now implies that in the interval $0 \leq x \leq x_{c0}$ there exist values of $x$ such that
\begin{equation}
K_{\eta}(x) \geq K_{\eta M}(x_{c0}) = { J \over 2 x_{c0}^2},
\label{KetaJ}
\end{equation}
subject to a 10 percent uncertainty.
The persistence of magnetic correlations with doping is sometimes viewed as indicating their strength and therefore favoring phase separation. The constraint (\ref{KetaJ}), however, suggests otherwise.

The identification of $x_{c0}$ is the subject of the rest of this paper.
This parameter is difficult to pinpoint, but the focus on finding it constitutes a new and, presumably, productive approach to the subject of phase separation in cuprates. The effectiveness of this approach is related to the fact that $K_{\eta M}$ depends very steeply on $x_{c0}$. The readers should appreciate that $K_{\eta M}$ for $x_{c0} = 0.14$ is 4 times larger than for $x_{c0} = 0.28$ and 4 times smaller than for $x_{c0} = 0.07$. Therefore, as long as other contributions to the curvature are known within  factor of two, and $x_{c0}$ is known with the accuracy of $\pm 0.05$, one can form a good judgment about the chances of phase separation.

Both experiments and numerical studies seem to agree that there are no noticeable AF correlations above $x=0.5$.
Further constraining $x_{c0}$ requires a more detailed analysis of the experimental phenomenology and theoretical scenarios.
This will be done in Sections~\ref{exper} and \ref{theor} respectively with the end result that both of these analyses point to $x_{c0} \approx 0.14$.

\subsubsection{Combining the estimates}
\label{combine}

\begin{figure} \setlength{\unitlength}{0.1cm}
%=======================================================================

\begin{picture}(100, 78) 
{ 
\put(5, 0){ \epsfxsize= 3in \epsfbox{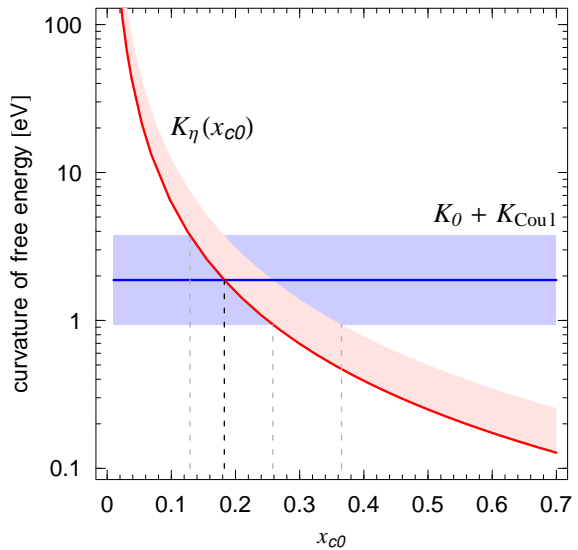} }
}
\end{picture} 
%============== 
\caption{ (Color online) Estimates for negative ($K_{\eta}$) and positive ($K_0 + K_{\hbox{\small Coul}}$) contributions to energy curvature per one in-plane Cu. Negative curvature is plotted as a function of unknown critical concentration $x_{c0}$. Solid lines represent the estimates by formulas (\ref{KetaJ}), (\ref{K0A}) and (\ref{KCoul}) with the numbers given in the text. Shaded areas around the lines cover the regions of the factor-of-two uncertainty for the above estimates.
} 
\label{fig-K} 
\end{figure}

Before trying to pin down the value of $x_{c0}$, it is useful to plot the bound on the value of $K_{\eta}$ as a function of $x_{c0}$ against the bounds on $K_0 + K_{\hbox{\small Coul}}$ thus comparing the two sides of condition (\ref{Ktot}) for phase separation. Such a plot is shown in Fig.~\ref{fig-K}.

%%%%%%%%%%%%%%%%%%%%%%%%%%%%%%%%%%%%%%%%%%%%%%%%%%%%%%%%%%%%%%%%%%%

% %%%%%%%%%%%%%%%%%%%%%%%%%%%%%%%%%%%%%%%%%%%%%%%%%%%%%%%%%%%%%%%%%

The horizontal line corresponds to the value of $K_0 + K_{\hbox{\small Coul}}$ estimated according to formulas (\ref{K0A}) and (\ref{KCoul}) with the following values of parameters: $a = 4 \ \hbox{\AA}$, $f^s_0 = 6$, $m^* = 4 m_e$, $R_s = 6\ \hbox{\AA}$, $\epsilon = 30$. According to this estimate $K_0 = 1.35 \ $eV  and $K_{\hbox{\small Coul}}= 0.57 \ $eV [both per in-plane Cu].  The shadowed stripe around the above line covers the region of the factor-of-two uncertainty in the estimated value.

The line corresponding to $K_{\eta} (x_{c0})$ is obtained using $K_{\eta M}(x_{c0})$ given by Eq.(\ref{KetaJ}) with $J=125 \ $meV. The shaded stripe above this line also represents the region of the factor-of-two uncertainty. This region does not spread below the line, because of the nature of constraint (\ref{KetaJ}). (Here the ten per cent uncertainty associated with condition (\ref{Fc}) and the small uncertainty in the knowledge of $J$ are neglected.)

The qualitative interpretation of Fig.~\ref{fig-K} can be phrased as follows:

(i) if $x_{c0} \leq 0.12$, then the system almost certainly phase separates;

(ii) if $0.12 < x_{c0} \leq 0.18$, then the system likely phase separates;

(iii) if $0.18 < x_{c0} \leq 0.26$, then the chances of phase separation are about 50 per cent;

(iv) for $0.26 < x_{c0} \leq 0.36$, the chances of phase separation are still significant but less then 50 per cent; 

(v) finally, for $0.36 < x_{c0}$, phase separation is increasingly unlikely.

\subsection{Constraint on $x_{c0}$ from experiments}
\label{exper}

Recent experiments by Wakimoto et al. \cite{Wakimoto-07} show that the intensity of AF correlations integrated in the frequency range 0---100~meV in La$_{2-x}$Sr$_x$CuO$_4$ at $x= 0.3$ is about 10 per cent of that in La$_{2-x}$Ba$_x$CuO$_4$ at $x=0.125$, which, in turn, should be significantly smaller than the AF intensity in the parent compound LaCuO$_4$ corresponding to $x=0$.  Similar results but with a smaller frequency integration range, 0---50meV, were also reported previously for YBa$_2$Cu$_3$O$_{6+x}$ by Bourges\cite{Bourges-00}. It thus appears that $x_{c0} \approx 0.3$ would satisfy condition (\ref{Fc}) defining the critical crossover concentration.

However,  $x_{c0}$ can also be significantly smaller than 0.3. If phase separation actually takes place in cuprates, then the experimental magnetic signal can come from hole-poor regions of the phase separated state. All we can conclude then is that $x_2 \approx 0.3$, but the maximum critical concentration $x_{c0}$ for the homogeneous state can be significantly smaller than $x_2$. [Here $x_2$ is the right end of the right metastable region defined in Section~\ref{free} and on Fig.~\ref{fig-F}(b).]

\begin{figure} \setlength{\unitlength}{0.1cm}
%=======================================================================

\begin{picture}(100, 62) 
{ 
\put(0, 0){  \epsfxsize= 3.2in \epsfbox{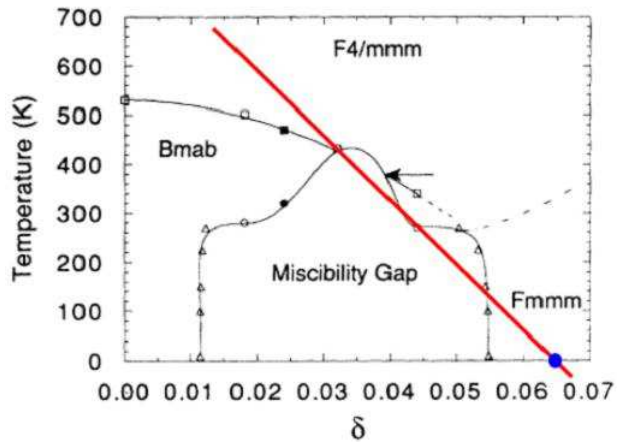} }
}
\end{picture}
%============== 
\caption{ (Color online) Experimental data points for the phase separation diagram of LaCuO$_{4+\delta}$ from Ref.~\cite{Radaelli-94} with solid straight line superimposed on the top of it to extrapolate to the right boundary of the phase separation region for the hypothetical case of never-frozen oxygen motion.
The nominal charge carrier concentration is $x = 2 \delta$.
} 
\label{fig-LCO} 
\end{figure}

We propose to further constrain $x_{c0}$ using the phenomenology of phase separation in LaCuO$_{4+\delta}$ (Fig.~\ref{fig-LCO}), and the fact  that for a phase diagram like the one shown in Fig.~\ref{fig-Tcx}, the temperature-dependent critical concentration, which we denote here as $x_c(T)$ should be close to the  right boundary between spinodal (locally unstable) region and metastable (locally stable) region. [{\it Note:} $x_{c0} \equiv x_c(0)$]. As shown in Appendix~\ref{classification}(b), $x_c(T)$ simply coincides with the spinodal boundary in the case of second order phase transitions.

Oxygen intercalated LaCuO$_{4+\delta}$ is an atypical cuprate family, where macroscopic phase separation actually takes place, supposedly because intercalated oxygen atoms remain mobile between CuO$_2$ planes above temperatures about 200K.  The redistribution of intercalated oxygen can thus screen the uncompensated charge of electronic inhomogeneities. A popular piece of the experimental phenomenology of phase separation in LaCuO$_{4+\delta}$ is shown in Fig.~\ref{fig-LCO}. The horizontal axis of Fig.~\ref{fig-LCO} represents the value of $\delta$ in LaCuO$_{4+\delta}$. It is tentatively assumed to be related to the charge carrier concentration as $x = 2 \delta$.

The energetics of intercalated oxygen can be a large part of the energy balance behind the phase separation in LaCuO$_{4+\delta}$. However, it is clear from our estimate in section~\ref{combine}, that the contribution to the phase separation energy balance from AF correlations in CuO$_2$ planes should also be large. When several large energy terms closely compete for and  against phase separation and the overall energy gain resulting from phase separation is small, then it is likely that the spinodal region ends where one of the large competing terms suddenly becomes small. In the present case, we know that the energy of AF correlations decreases steeply in the relevant doping region. Therefore, it is natural to expect that the spinodal region in LaCuO$_{4+\delta}$ ends where the AF correlations either disappear or are drastically reduced.

%%%%%%%%%%%%%%%%%%%%%%%%%%%%%%%%%%%%%%%%%%%%%%%%%%%%%%%%%%%%%%%%%%%

% %%%%%%%%%%%%%%%%%%%%%%%%%%%%%%%%%%%%%%%%%%%%%%%%%%%%%%%%%%%%%%%%%

It is not entirely straightforward to extract the boundary of spinodal region from experiments. For infinitely slow cooling, one expects that the system starts phase separating once the value of $x_2(T)$ (the right boundary of metastable region) becomes equal to the externally set doping level. However, for a finite cooling rate, the onset of phase separation may happen at lower temperatures, because it requires activation over an energy barrier between metastable homogeneous state and phase-separated state. One can only be certain, that, even if the cooling rate is too fast (but the oxygen ions remain mobile), phase separation should start once the charge carrier concentration reaches the right spinodal boundary, which we associated with the critical concentration $x_c(T)$. In other words:
\begin{equation}
x_c(T) \leq x_{exp}(T),
\label{xcxexp}
\end{equation}
where $x_{exp}(T)$ is the right boundary of the phase separation range observable experimentally {\it at temperatures, where intercalated oxygen ions are still mobile}.  The loss of oxygen mobility manifests itself in the freezing of the experimentally observable boundaries of the phase separation region, i.e. these boundaries in the axes of Fig.~\ref{fig-LCO} become vertical.   One can then see that in Fig.~\ref{fig-LCO} there are only two useful experimental points for the right end of the phase separation region --- those through which the straight line is drawn. This straight line is a crude extrapolation aiming at finding the right boundary of the phase separation range for the hypothetical case of never-frozen oxygen ions. This line crosses the horizontal axis at $\delta = 0.065$, which, according to the assumption $x = 2 \delta$, implies $x_{exp}(0) =  0.13$. Given inequality (\ref{xcxexp}) and recalling that $x_{co} \equiv x_c(0)$, we then obtain
\begin{equation}
x_{c0} \leq 0.13 \ \ .
\label{xc013}
\end{equation}
It had been reported\cite{Li-96} that, in fact, $x < 2 \delta$. This, however, only strengthens the case for inequality (\ref{xc013}).

To summarize this subsection, we have discussed the direct observational constraint\cite{Wakimoto-07} $x_{c0}< 0.3$ and an indirect experimental evidence (Fig.~\ref{fig-LCO}) for $x_{c0}< 0.13$. In terms of subsection~\ref{combine}, $x_{c0}= 0.3$ implies that the chances of phase separation are significant but less than 50 per cent, while $x_{c0}< 0.13$ indicates that phase separation is likely, i.e. its chances are significantly above 50 per cent.

\subsection{Theoretical discussion of microscopic factors}
\label{theor}

The studies of the $t-J$ and Hubbard models for the values of parameters relevant to the  cuprates\cite{Emery-90,Dagotto-94,Nagaev-95,Hellberg-97,Becca-00,Ivanov-04,Aichhorn-05,Aichhorn-06,Lugas-06,Macridin-06,Eckstein-07,Kocharian-07} agree that these models are close to the threshold of phase separation but do not reach the consensus on whether phase separation actually takes place or not.  However, the common expectation is  that even if the $t-J$ or Hubbard model solutions are slightly over the phase separation threshold, the  in-plane Coulomb repulsion between charge carriers (not included in the models but present in real materials) would easily suppress phase separation. Ivanov\cite{Ivanov-04} has reported numerical results directly supporting this expectation.

The above expectation appears to be in conflict with the conclusion of the preceeding subsection derived from the analysis of experiments. Below we further suggest that this expectation can be misleading on purely theoretical grounds.
We argue that the omission of the ``medium-range" Coulomb repulsion between the charge carriers separated by 1-3 lattice periods leads to a larger value of the critical concentration $x_{c0}$, where AF correlations vanish. This, in turn, weakens the tendency towards phase separation.

\

\begin{figure} \setlength{\unitlength}{0.1cm}
%=======================================================================

\begin{picture}(100, 65) 
{ 
\put(5, 0){ \epsfxsize= 3in \epsfbox{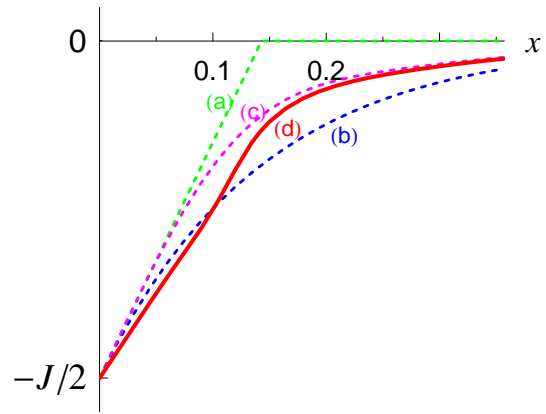} }
}
\end{picture} 
%============== 
\caption{ (Color online) Illustrations for various scenarios for the doping evolution of the AF exchange energy in cuprates: a) independent charge carriers; b) correlated charge carriers in AF background; c) medium-range Coulomb repulsion between charge carriers added; d) Coulomb potential of dopant atoms added. See the discussion in the text.
} 
\label{fig-theor} 
\end{figure}

We introduce the effect of the medium-range Coulomb interactions on the doping dependence of AF correlations in several steps as illustrated in Fig.~\ref{fig-theor}. We start from the half-filled state, where the exchange energy per spin is about $J/2$, and then estimate $x_{c0}$ as the concentration of holes necessary to destroy this energy. Although $F_{\eta}(x)$ includes indirect effects of AF correlations in addition to just AF exchange energy, this function should become zero, when AF correlations disappear, which, in turn, happens simultaneously with AF exchange energy falling to zero.

%%%%%%%%%%%%%%%%%%%%%%%%%%%%%%%%%%%%%%%%%%%%%%%%%%%%%%%%%%%%%%%%%%%

% %%%%%%%%%%%%%%%%%%%%%%%%%%%%%%%%%%%%%%%%%%%%%%%%%%%%%%%%%%%%%%%%%

First we consider the approximation of ``independent charge carriers'' [line (a) in Fig.~\ref{fig-theor}], according to which, each hole independently destroys AF correlations in the region around itself. The AF energy cost of a doped hole can be estimated\cite{Shraiman-88} using the typical length of two lattice periods for the string of broken AF bonds, which are left behind by the hole added to the half-filled t-J model with $t/J = 3$. This means that 1 hole creates approximately 7 ferromagnetic bonds at energy cost $J/2$ each, i.e. it destroys AF energy about ${7 \over 2} J$. This simple estimate should be reasonable for a broader class of situations beyond the string picture in $t-J$ model.

In the above approximation, the AF energy disappears at the doping concentration determined by condition:
\begin{equation}
{7 \over 2} J x_{c0} = {1 \over 2} J,
\label{Jcond} 
\end{equation}
which means $x_{c0} \approx 0.14$.

The above estimate of the AF exchange energy per hole should hold for very small doping concentrations. 
As Step 2 of our analysis, we take into account the fact that, at larger concentrations, the holes would tend to hop onto the sites with AF bonds broken by a nearby hole. The AF exchange energy cost per hole is, therefore, reduced as the hole concentration increases, following line (b) sketched in Fig.~\ref{fig-theor}. This sketch presumably reflects what happens in the t-J and Hubbard models and explains why $x_{c0}$ associated with those models can easily increase to the values 0.4-0.5, which, in turn, makes the onset of phase separation uncertain.

At the next step, we introduce the medium-range Coulomb repulsion between charge carriers. This repulsion prevents one hole from  taking advantage of  the AF bonds broken by another hole. Therefore, each hole again destroys more AF bonds.
It is this factor, that may have been missing in the numerical work of Ivanov\cite{Ivanov-04}, who found that the nearest neighbor Coulomb repulsion suppressed phase separation. In Ref.~\cite{Ivanov-04}, however, a variational ansatz was used, which did not suppress the chances of holes to occupy adjacent lattice sites, and, therefore, not surprisingly, the resulting energy balance opposed phase separation.

One can estimate the Coulomb interaction constants between holes occupying adjacent lattice sites as 700~meV. This estimate uses the high frequency dielectric constant $\epsilon \approx 5$. The high frequency value is taken, because the lifetime of the region of broken AF bonds can be estimated as  $1/J$, which is faster than the  timescale of lattice vibrations responsible for low frequency Coulomb screening. (We assume the Debye frequency $ 60$~meV$/\hbar$.)

One should also consider the possibility of holes forming long-living pairs or larger clusters. In this case, lattice would have time to respond, which means that it would additionally screen the interaction between holes, and thereby reduce the dielectric constant to $\epsilon \approx 30$\cite{Kastner-98}.
However, even in the latter case, the repulsion between holes occupying nearest neighbor sites remains significant --- about 100~meV.

Once the medium-range Coulomb repulsion is taken into consideration, the plot of exchange energy as a function of doping concentration should pass somewhere between lines (a) and (b) in Fig.~\ref{fig-theor}, which gives line (c). At lower concentrations, it is closer to the independent hole approximation, i.e. line (a), and at higher concentrations, when charge carriers have no choice but to approach each other, the plot should be closer to line (b).
Thus, the effect of the medium range Coulomb repulsion is to make the AF crossover sharper, i.e. the AF negative curvature in the crossover region around $x \approx 0.14$  should become larger.

The next level of approximation is to introduce the Coulomb potential of dopant ions --- line (d) in Fig.~\ref{fig-theor}. At low doping concentrations, this potential should localize holes around dopant ions and thereby reduce the radius of the region around the hole, where AF correlations are destroyed. Therefore, the low doping part of the AF energy vs. doping concentration plot should have smaller slope than the one in the preceeding step. As the doping concentration increases, the Coulomb potential wells around the dopant ions start overlapping stronger, and therefore, the result of the previous step is recovered. As is obvious from Fig.~\ref{fig-theor} such a correction further increases the AF negative curvature in the region around $x \approx 0.14$. Quantitatively, however, this correction depends on the localization radius around dopant ions and may be minor, if this radius is larger than two lattice periods. The latter appears to be the case for the cuprates\cite{Kastner-98}.

Finally, the Coulomb interaction would also play an important role in the energy balance determining what kind of inhomogeneous patterns emerge if the homogeneous state becomes unstable, but this is the subject beyond the scope of the present work. As mentioned in subsection~\ref{Coulomb}, the above instability can lead, among others, to stripes\cite{Tranquada-95}, checkerboards(e.g. \cite{Fine-hitc-prb04,Fine-vortices-prb07}), or the charge imbalance between adjacent CuO$_2$ planes.

\section{Conclusions}

We have demonstrated that focusing on the question of how the energy associated with magnetic or other type of correlations evolves with doping, leads to a useful insight into the factors controlling phase separation in the vicinity of phase transitions and crossovers. We have introduced a quantitative constraint on the negative curvature contribution to the free energy and used this constraint to evaluate the chances of electronic phase separation in the cuprates.

Our analysis of the antiferromagnetic crossover in the cuprates has led us to  conclude that these materials are realistically close to the phase separation threshold, and the chances that they generically exhibit some form of phase separation within the superconducting doping range are quantitatively significant. In particular, the lasagna phase separation scenario with  macroscopic regions of opposite charge lying on the top of each other in adjacent CuO$_2$ planes appears to be quite viable and creates additional worry  for the interpretation of many experiments. Cuprates also appear to be closer to the phase separation threshold than the expectations based on the Hubbard and $t-J$ models suggest. This may be related to the fact that the above models neglect Coulomb interaction between charge carriers separated by 1-3 lattice sites.

The proximity to the phase separation threshold on either side, rather than the presence or the absence of phase separation as such, may be the distinguishing characteristics of the superconducting cuprates. The resulting ``softness" of the electronic liquid should then lead to significant fluctuations of charge density and electric field at long wavelengths and relatively low frequencies\cite{Leggett-99}.

This work was supported in part by the National Science Foundation
through DMR-0404781.

\

\

\

\

\appendix

\section{Classification of homogeneous scenarios behind phase separation}
\label{classification}

\subsubsection{First order phase transitions}
\label{first-order}

At a first order phase transition, the system switches from one minimum of free energy to another, not continuously connected to the first one in the space of macroscopic variables of the system. The values of free energy of the minima should have different dependencies on the carrier concentration. The switching between the two minima takes place at the critical carrier concentration, where the two free energies coincide (see Fig.~\ref{fig-cusp}).  As obvious from Fig.~\ref{fig-cusp}, this guarantees that the resulting concentration dependence of the minimal free energy has an upward cusp at $x = x_{c0}$. The upward cusp implies infinite negative second derivative at $x = x_{c0}$, which, in turn, guarantees that the system phase separates around $x_{c0}$ [at least on nanoscale], no matter what the value of Coulomb interaction is.

Since the observation of phase separation does not immediately imply that the underlying homogeneous transition would be of the first order, one can make a compelling case in favor of such a transition, only when none of the two phases can emerge from the other as a result of spontaneous symmetry breaking: e.g., for antiferromagnetic---ferromagnetic transition in the context of colossal magnetoresistance manganites, and for the low temperature orthorhombic---low temperature tetragonal (LTO---LTT) transition in cuprates and nickelates. However, even in the above two examples, the homogeneous transition can be split into two closely spaced second order transitions resulting in a ``rounded'' cusp
in the concentration dependence of free energy.

\subsubsection{Second order phase transitions}
\label{second-order}

We start from the standard Landau expansion for a second order phase transition in terms of an order parameter having absolute value $\eta$ (e.g. the absolute value of staggered magnetization).  The transformational properties of this order parameter under time reversal are such that, in the absence of external field, only even powers of $\eta$ enter the expansion for the free energy, i.e.
\begin{equation}
F_{\eta} = A \eta^2 + B \eta^4,
\label{Feta}
\end{equation}
where $A$ and $B$ are two expansion coefficients.
The onset of the  phase transition is determined by the sign of coefficient $A$, which can be parameterized as
$A = \alpha (T - T_c)$,
where $\alpha$ is a positive constant, $T$ is temperature, and $T_c$ is the transition temperature.
At $T<T_c$, $A$ is negative and hence the minimum of free energy is reached at
$\eta = \sqrt{\alpha (T_c - T) \over 2 B}$.
The minimized free energy is then
\begin{equation}
F_{\eta} = - {\alpha^2 (T_c - T)^2 \over 4 B}.
\label{Fmin}
\end{equation}

Now we consider the evolution of parameters in Eq.(\ref{Fmin}) as a function of carrier concentration as relevant to the phase diagram shown in Fig.~\ref{fig-Tcx}.
The ratio ${\alpha^2 \over 4 B}$ may depend on the concentration $x$, but near (below) the critical doping concentration $x_{c0}$, the key factor is the dependence of $T_c$ on $x$, which we parameterize in the vicinity of $x_{c0}$ as
\begin{equation}
T_c = \lambda (x_{c0} - x),
\label{Tc}
\end{equation}
where $\lambda$ is a slope parameter.
Any alternative parameterization of the form \mbox{$T_c = \lambda (x_{c0} - x)^{\zeta}$} with $\zeta >1$ or $0 \leq \zeta \leq 1$ would be more favorable to phase separation (see Fig.~\ref{fig-alt} and the related discussion in subsection~\ref{rigorous}).
Substituting Eq.(\ref{Tc}) into Eq.(\ref{Fmin}), we obtain
\begin{equation}
F_{\eta} =  - K_{\eta} [x_c(T) - x]^2,
\label{Fetax} 
\end{equation}
where $K_{\eta}$ the {\it negative} curvature given by
\begin{equation}
K_{\eta} = {\lambda^2 \alpha^2 \over 4 B},
\label{Keta}
\end{equation}
and
\begin{equation}
x_c(T) = x_{c0} - {T \over \lambda}.
\label{xc}
\end{equation}
Important for the analysis of cuprates is the property that, if the negative curvature (\ref{Keta}) outweighs the positive curvature of the rest of the free energy, then one of the ends of the miscibility gap would coincide with $x_c(T)$.

The value of $K_{\eta}$ given by Eq.(\ref{Keta}) can now be estimated as follows: 

The slope of the $T_c$-vs.-$x$ line around $x=x_{c0}$ can be crudely approximated as
\begin{equation}
\lambda = {T_{c0} \over x_{c0}},
\label{gamma} 
\end{equation}
where $T_{c0}$ is the critical temperature at $x=0$. The value of $\alpha^2 / (4B)$ can be estimated by equating the Landau expansion result and the approximate microscopic value of the gain in the ordering energy $F_{\eta}(x)$ at $x=0$, $T=0$:
\begin{equation}
F_{\eta 0} = {\alpha^2 \over 4B} T_{c0}^2 
\label{alphaB} 
\end{equation}
The value of $F_{\eta 0}$ is assumed to be known either from experiments or from direct microscopic considerations.
Thus 
\begin{equation}
K_{\eta} = {F_{\eta 0} \over x_{c0}^2}.
\label{curv}
\end{equation}

Despite the crudeness of the assumptions that led to Eq.(\ref{curv}), the right-hand-side of this equation is, in fact, a rigorous constraint on the negative curvature for a broad class of situations.  It is formulated  in subsection~\ref{rigorous} and derived in Appendix~\ref{constraints}.

\subsubsection{Crossovers}

\label{crossovers}

Crossovers constitute, perhaps, the most general case, where neither of the derivatives $F_{\eta}(x)$ exhibits a jump (see Fig.~\ref{fig-cross}). For the purposes of this work, phase transitions of orders higher than two can also be treated as crossovers. Like second order phase transitions the crossovers must lead to the regions of the negative curvature of $F_{\eta}(x)$. One can make a good estimate of this curvature by approximating the crossover behavior of $F_{\eta}(x)$ by a suitable second order phase transition curve. This procedure is put on a firm foundation in Section~\ref{rigorous} and Appendix~\ref{constraints} .

\section{Derivations of the constraints on the curvature.}
\label{constraints}

In this Appendix, we present the rigorous solutions of problems labeled as B, B1 and B2, which are equivalent, in respective order, to problems A, A1, and A2 formulated in Section~\ref{rigorous}.
While the problems A, A1 and A2 are of physical interest to us, the alternative formulation allows us to avoid many unnecessary minus signs and factors 1/2 in the solution. Everywhere below, variables without subscripts, like $F(x)$, $F^{\prime}(x)$ and $F^{\prime \prime}(x)$, are the functions of the argument in the parentheses, whereas subscripted variables like $F_1$, $F_2$, $F^{\prime}_1$, $F^{\prime}_2$, $F^{\prime}_{max}$, $F^{\prime \prime}_{max}$, etc., are only the numbers and, if followed by parentheses should be multiplied by the expression in the parentheses. As in Section~\ref{rigorous}, prime and double prime superscripts denote the first and the second derivatives respectively.

\begin{figure} \setlength{\unitlength}{0.1cm}
%=======================================================================

\begin{picture}(100, 65) 
{ 
\put(5, 0){ \epsfxsize= 3in \epsfbox{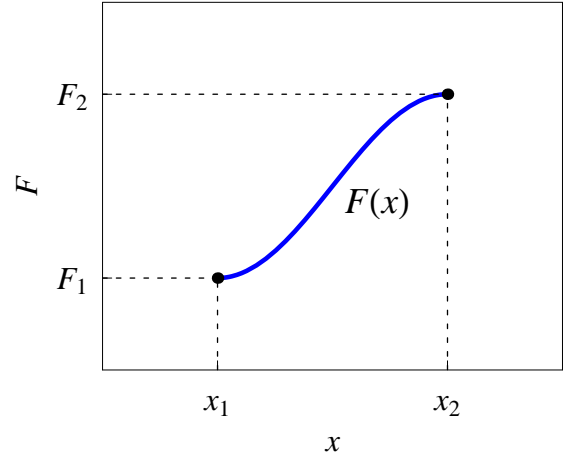} }
}
\end{picture} 
%============== 
\caption{ (Color online) Illustration for the boundary conditions of problem B.
} 
\label{fig-problB} 
\end{figure}

{\it Problem B}: 
Here  we consider all paths leading from $x=x_1$ to $x=x_2 > x_1$, such that $F(x_1) = F_1$, \mbox{$F^{\prime}(x_1-0) = 0$}  and $F(x_2)= F_2 > F_1$, and the question is: What is the {\it minimum} possible value of the maximum {\it positive} second derivative 
(max$[F^{\prime \prime} (x)]$) on a path leading from $F(x_1)$ to $F(x_2)$? See Fig.~\ref{fig-problB}

%%%%%%%%%%%%%%%%%%%%%%%%%%%%%%%%%%%%%%%%%%%%%%%%%%%%%%%%%%%%%%%%%%%

% %%%%%%%%%%%%%%%%%%%%%%%%%%%%%%%%%%%%%%%%%%%%%%%%%%%%%%%%%%%%%%%%%

Functions $F(x)$ need not be analytical. Here, the infinite positive second derivative $F^{\prime \prime}(x)$ would disqualify the path, while infinite negative second derivative may be present in the optimal path.
With the above reservations, we will formally treat paths $F(x)$ as if they were analytical functions.

{\it Solution of Problem B.}

Let us denote the maximum values of $F^{\prime}(x)$ and $F^{\prime \prime}(x)$ in the interval $[x_1, x_2]$ as $F^{\prime}_{max}$ and $F^{\prime \prime}_{max}$, respectively and the value of $x$, where $F^{\prime}(x)$ reaches its maximum, as $x_m$.
We now express $F_2$ as follows
\begin{equation}
F_2 = F_1 + \int_{x_1}^{x_m} F^{\prime}(x) \ dx + \int_{x_m}^{x_2} F^{\prime}(x) \ dx .
\label{F2}
\end{equation}
Let us now denote the maximum value of $F^{\prime \prime}(x)$ within the first interval 
$[x_1, x_m]$ as $\tilde{F}^{\prime \prime}_{max}$.  Since $F^{\prime}(x_1) = 0$, the value of 
$F^{\prime}(x)$ within the above interval can be constrained as follows
\begin{equation}
F^{\prime}(x) = \int_{x_1}^{x} F^{\prime \prime}(x^{\prime}) \ dx^{\prime} \leq 
\tilde{F}^{\prime \prime}_{max} (x-x_1).
\label{Fpr}
\end{equation}
Inequality (\ref{Fpr}) has two consequences
\begin{equation}
F^{\prime}_{max} \leq \tilde{F}^{\prime \prime}_{max} (x_m - x_1)
\label{Fprmax1}
\end{equation}
and
\begin{equation}
\int_{x_1}^{x_m} F^{\prime}(x) \ dx \leq 
{1 \over 2} \tilde{F}^{\prime \prime}_{max} (x_m - x_1)^2 .
\label{intFpr1}
\end{equation}

For the second integral in Eq.(\ref{F2}), we use the bound
\begin{equation}
\int_{x_m}^{x_2} F^{\prime}(x) \ dx \leq F^{\prime}_{max} (x_2 - x_m)
\label{intFpr2}
\end{equation}
Now, substituting the constraint (\ref{Fprmax1}) for $F^{\prime}_{max}$ into inequality (\ref{intFpr2}) and combining the result with Eq.(\ref{F2}) and  inequality (\ref{intFpr1}), we obtain
\begin{equation}
F_2 \leq F_1 + {1 \over 2} \tilde{F}^{\prime \prime}_{max} 
\left[ (x_2 - x_1)^2 -(x_2 - x_m)^2 \right]
\label{intF2a}
\end{equation}
The expression in the square brackets above is always positive in the interval $[x_1, x_2]$
and has maximum value, when $x_m = x_2$. Taking this into account together with the fact that
$F^{\prime \prime}_{max} \geq \tilde{F}^{\prime \prime}_{max}$, we, finally, obtain
\begin{equation}
F^{\prime \prime}_{max} \geq {2 (F_2 - F_1) \over (x_2 - x_1)^2}  .
\label{Fdprmax2}
\end{equation}
The right-hand side of the above inequality, actually, represents a possible value of the maximum curvature, which corresponds to parabola
\begin{equation}
F(x) = F_1 + {F_2 - F_1 \over (x_2 - x_1)^2} (x - x_1)^2.
\label{parabola}
\end{equation}
Hence the right-hand side of inequality (\ref{Fdprmax2}), is the minimum possible value of $F^{\prime \prime}_{max}$. In terms of Problem A, this result implies formula (\ref{minKmax1}). The A-problem counterpart of parabola (\ref{parabola}) is shown in Fig.\ref{fig-alt}.

\

Now we generalize Problem B to {\it Problem B1} (equivalent to Problem A1 from Section~\ref{rigorous}). Problem B1 has all conditions of Problem B and, in addition:
$ F^{\prime}(x_1) = F^{\prime}_1$. All other notations will be the same as in the solution of Problem B.

{\it Solution of Problem B1}

One can reduce Problem B1 for $F(x)$ to Problem B for function
\begin{equation}
G(x) = F(x) - F^{\prime}_1 (x-x_1).
\label{G}
\end{equation}
In this case,  $G(x_1)\equiv G_1 = F_1$, and
$G(x_2)\equiv G_2 = F_2 - F^{\prime}_1 (x_2-x_1)$
The curvatures of $G(x)$ and $F(x)$ are, obviously, the same everywhere .
If $G_2 > G_1$, or, equivalently,
\begin{equation}
F^{\prime}_1 > {F_2 - F_1 \over x_2 - x_1},
\label{Fp1}
\end{equation}
then inequality (\ref{Fdprmax2}) translates into
\begin{equation}
F^{\prime \prime}_{max} \geq {2 (G_2 - G_1) \over (x_2 - x_1)^2} =  
{2 [F_2 - F_1 - F^{\prime}_1 (x_2-x_1)] \over (x_2 - x_1)^2} .
\label{Fdprmax3}
\end{equation}
If condition (\ref{Fp1}) is not satisfied, then there are paths between $x_1$ and $x_2$,
which have no regions of positive curvature. Inequality (\ref{Fdprmax3}) is equivalent to Eq.(\ref{minKmax2}) for Problem A1.

Let us now turn to {\it Problem B2}, which has condition $F^{\prime}(x_2) = F^{\prime}_2$ in addition to all other conditions of Problem B1.

Let us further assume that $F^{\prime}(x_1) = 0$. (If $F^{\prime}(x_1) \neq 0$, then one can perform transformation (\ref{G}) and bring the problem to the condition desired.)

We first consider the  situation (Case I), when 
\begin{equation}
F^{\prime}_2 \leq {2 (F_2 - F_1) \over x_2 - x_1}.
\label{Fp2}
\end{equation}
The right-hand side of the above inequality is the first derivative at $x = x_2$ of the optimal parabolic path (\ref{parabola}) achieving the minimum of $F^{\prime \prime}_{max}$ in the absence of constraint  on $F^{\prime}(x_2)$.

Now we make the following two remarks:

1. The minimum of $F^{\prime \prime}_{max}$ in the present case (with constraint on $F^{\prime}(x_2)$) cannot be smaller than the minimum obtained for the case without the constraint on $F^{\prime}(x_2)$. Therefore, if we construct a path, whose  
$F^{\prime \prime}_{max}$ can be made arbitrarily close to the minimum $F^{\prime \prime}_{max}$ for the constraint-free case, then the constraint-free minimum will simultaneously be the minimum for the present case.

2. Possible paths can have points of infinite negative second derivative (see the earlier discussion).

The example anticipated in the first remark, indeed exists. The required path consists of a parabola, which approaches arbitrarily close from above to the optimal parabola (\ref{parabola}) in the case without constraint on $F^{\prime}(x_2)$. When the former parabola crosses the straight line described by function $F(x) = F_2 + F^{\prime}_2 (x - x_2)$, the path just switches to that straight line thus exhibiting infinite negative second derivative. The maximum positive second derivative on such a path is that of its parabolic part, which, therefore, can have value arbitrarily  close to that of the optimal parabola (\ref{parabola}). Thus Case I of the present problem with $F^{\prime}(x_1) = 0$ results in condition $\ref{Fdprmax2}$.

Now we turn to Case II, corresponding to condition
\begin{equation}
F^{\prime}_2 > {2 (F_2 - F_1) \over x_2 - x_1}.
\label{Fp2a}
\end{equation}
In this case, the straightforward generalization of the solution for Case I would not work, because the switching between the parabola and the straight line would have to exhibit infinite positive second derivative, which would be contrary to our task of minimizing 
$F^{\prime \prime}_{max}$. 

Instead we define a new function
\begin{equation}
H(x) = F(x) - F^{\prime}_2 \ (x-x_2),
\label{H}
\end{equation}
which everywhere has the same second derivative as $F(x)$. The new function is constrained by the following conditions:
\begin{equation}
H(x_1) \equiv H_1  = F_1 + F^{\prime}_2 \ (x_2-x_1),
\label{H1}
\end{equation}
\begin{equation}
H^{\prime}(x_1) \equiv H^{\prime}_1  = - F^{\prime}_2 ,
\label{Hp1}
\end{equation}
\begin{equation}
H(x_2) \equiv H_2  = F_2 ,
\label{H2}
\end{equation}
\begin{equation}
H^{\prime}(x_2) \equiv H^{\prime}_2  = 0 .
\label{Hp2}
\end{equation}
Conceptually, the problem for $H(x)$ is a mirror reflection of our original problem for $F(x)$, and, moreover, Case II of the problem for $F(x)$ corresponds to
Case I for $H(x)$. The latter is defined by the condition
\begin{equation}
|H^{\prime}_1 | \leq {2 (H_1 - H_2) \over x_2 - x_1}.
\label{Hp1a}
\end{equation}
which is, indeed, fulfilled given  inequality (\ref{Fp2a}) and Eqs.(\ref{H1}-\ref{H2}). This means that the path minimizing the maximum positive curvature for both $H(x)$ and $F(x)$ is a parabola, which matches the constraint on the first derivative at $x = x_2$ and switches to a straight line near $x= x_1$. It is characterized by
\begin{equation}
F^{\prime \prime}_{max}  =  
{2 (F_1 - F_2 + F^{\prime}_2 (x_2-x_1)) \over (x_2 - x_1)^2} .
\label{Fdprmax4}
\end{equation}

We conclude this Appendix by giving explicit expressions for the  results of Problem B2 for the case, when $F^{\prime}(x_1) \neq 0$ and $F^{\prime}(x_2) \neq 0$:

1. If $F^{\prime}(x_1) > {(F_2 - F_1) \over x_2 - x_1} $ and either there is no constraint on $F^{\prime}(x_2)$, or  $F^{\prime}(x_2) < {(F_2 - F_1) \over x_2 - x_1} $, then the paths between $x_1$ and $x_2$ can have no regions of positive curvature.

Otherwise:

2a. If  either $F^{\prime}(x_1) + F^{\prime}(x_2) \leq {2 (F_2 - F_1) \over x_2 - x_1} $ or $F^{\prime}(x_2)$ is unknown, then the minimum value of $F^{\prime \prime}_{max}$ is given by the right-hand side of inequality (\ref{Fdprmax3}).

2b. If $F^{\prime}(x_2)$ is known and 
$F^{\prime}(x_1) + F^{\prime}(x_2) > {2 (F_2 - F_1) \over x_2 - x_1} $,
then the minimum of $F^{\prime \prime}_{max}$ is given by Eq.(\ref{Fdprmax4}).

In both cases 2a and 2b, the minimal values of $F^{\prime \prime}_{max}$ correspond to parabolas passing through points $(x_1, F_1)$ and $(x_2, F_2)$. In case 2a, the optimal parabola matches the given value of $F^{\prime}(x_1)$, while in case 2b it matches
$F^{\prime}(x_2)$.

In the end of Section~\ref{rigorous}, the above results are reexpressed for the negative curvature in terms of Problem A2.

%\bibliography{spinod}

\begin{thebibliography}{45}
\bibitem[{cus()}]{cusp}
\bibinfo{note}{Less generic situations occur either, when $F_{\hbox{\small
  tot}} (x)$ exhibits an upward cusp in the interval between $x_1$ and $x_2$
  (first order phase transition), or, when it has downward cusps at $x_1$ or
  $x_2$ (the chemical potential crosses a gap in the density of states). In the
  former case, the spinodal region may be reduced to a single point. In the
  later case, one or two metastable regions may be reduced to a point.}
\end{thebibliography}

\end{document}